\documentclass[%
reprint,
nofootinbib,
 amsmath,amssymb,
 aps,
]{revtex4-1}

\usepackage{graphicx}
\usepackage{dcolumn}
\usepackage{bm}

\begin{document}


\title{Origin of the light cosmic ray component below the ankle}

\author{A.D.~Supanitsky}
\affiliation{Universidad de Buenos Aires. Facultad de Cs.~Exactas y Naturales. Buenos Aires, Argentina\\}
\affiliation{CONICET-Universidad de Buenos Aires. Instituto de Astronom{\'i}a y F{\'i}sica del Espacio (IAFE)
CC 67, Suc.~28, 1428 Buenos Aires, Argentina}
\email{supanitsky@iafe.uba.ar}

\author{A.~Cobos} 
\affiliation{Instituto de Tecnolog\'ias en Detecci\'on y Astropart\'iculas Mendoza (CNEA, CONICET, UNSAM),
Mendoza, Argentina.}

\author{A.~Etchegoyen}
\affiliation{Instituto de Tecnolog\'ias en Detecci\'on y Astropart\'iculas (CNEA, CONICET, UNSAM),
Centro At\'omico Constituyentes, San Martin, Buenos Aires, Argentina.}

\date{\today}

\begin{abstract}

The origin and nature of the ultrahigh energy cosmic rays remains a mystery. However, 
considerable progress has been achieved in past years due to observations performed by the Pierre 
Auger Observatory and Telescope Array. Above $10^{18}$ eV the observed energy spectrum presents 
two features: a hardening of the slope at $\sim 10^{18.6}$ eV, which is known as the ankle, and 
a suppression at $\sim 10^{19.6}$ eV. The composition inferred from the experimental data, 
interpreted by using the current high energy hadronic interaction models, seems to be light 
below the ankle, showing a trend to heavier nuclei for increasing values of the primary energy. 
Also, the anisotropy information is consistent with an extragalactic origin of this light 
component that would dominate the spectrum below the ankle. Therefore, the models that explain 
the ankle as the transition from the galactic and extragalactic components are disfavored by 
present data. Recently, it has been proposed that this light component originates from the 
photodisintegration of more energetic and heavier nuclei in the source environment. The 
formation of the ankle can also be explained by this mechanism. In this work we study in 
detail this general scenario but in the context of the central region of active galaxies. 
In this case, the cosmic rays are accelerated near the supermassive black hole present in 
the central region of these types of galaxies, and the photodisintegration of heavy nuclei 
takes place in the radiation field that surrounds the supermassive black hole.

\end{abstract}

\maketitle


\allowdisplaybreaks 

\section{Introduction}

The origin of the ultrahigh energy cosmic rays (UHECRs), i.e., with energies above $10^{18}$ eV,
is still unknown. The three main observables used to study their nature are the energy spectrum,
composition profile, and distribution of their arrival directions. In this energy range, 
these studies are carried out by detecting the atmospheric air showers initiated by the UHECR 
primaries that interact with molecules of the atmosphere. The most common detection systems 
include arrays of surface detectors, which allow reconstructing the lateral development of the 
showers by detecting secondary particles that reach the ground, and fluorescence telescopes, 
which are used to study the longitudinal development of the showers. The two observatories 
currently taking data are the Pierre Auger Observatory \cite{Auger:15}, situated in the southern
hemisphere in Malarg{\"u}e, Province of Mendoza, Argentina, and Telescope Array \cite{TA:03}, 
located in the northern hemisphere, in Utah, United States. Both observatories combine arrays of 
surface detectors with fluorescence telescopes.

The ankle in the UHECR flux has been observed by several experiments \cite{Patrignani:16}. 
The Pierre Auger Observatory observes this spectral feature at an energy 
$E_{\textrm{ankle}}=10^{(18.705 \pm 0.005)}$ eV \cite{AugerICRC:17}. At this point, the spectral 
index, assuming the differential flux to be given by a power law $J \propto E^{-\gamma}$, changes 
from $\gamma_1=-3.293 \pm 0.002$ below the ankle to $\gamma_2=-2.53 \pm 0.02$ 
above the ankle \cite{AugerICRC:17}. Similarly, Telescope Array observes the ankle at 
$E_{\textrm{ankle}}=10^{(18.71 \pm 0.02)}$ eV and reports a change in the spectral index from 
$\gamma_1=-3.246 \pm 0.005$ below to $\gamma_2=-2.66 \pm 0.03$ above the break \cite{IvanovTA:15}. 
The suppression of the flux is observed at $E_{\textrm{s}}=10^{(19.59 \pm 0.02)}$ eV in the case 
of Auger \cite{AugerICRC:17} and at $E_{\textrm{s}}=10^{(19.75 \pm 0.05)}$ eV in the case of 
Telescope Array \cite{IvanovTA:15}. Even though these two values have been obtained by fitting 
the respective energy spectrum with different functions, it can be seen that the suppression of 
the spectrum is observed by Auger at a smaller energy than Telescope Array. Also, the Auger 
spectrum takes smaller values than the ones corresponding to Telescope Array. The discrepancies 
between the two observations can be diminished by shifting the energy scales of both experiments 
within their systematic uncertainties. However, some differences are still present in the 
suppression region \cite{IvanovAugerTA:17}.

It is very well known that the most sensitive parameters to the nature of the primary particle 
are the muon content of the showers and the atmospheric depth of the shower maximum, $X_{max}$
(see, for instance, Ref.~\cite{Supanitsky:08}). The $X_{max}$ parameter can be reconstructed from 
the data taken by the fluorescence telescopes. The secondary charged particles of the showers 
interact with the nitrogen molecules of the atmosphere producing fluorescence light. Part of 
this light is detected by the telescopes that take data on clear and moonless nights. In this 
way, it is possible to observe the longitudinal development of the showers, which in turn may be
analyzed to infer the $X_{max}$ parameter. As mentioned before, this technique is employed 
by both Auger and Telescope Array. 

The composition analyses are performed by comparing experimental data with simulations of the showers.
These simulations are subject to large systematic uncertainties because they are based on high energy 
hadronic interaction models that extrapolate low energy accelerator data to the highest energies. 
Recently, the high energy hadronic interaction models more frequently used in the literature have been 
updated by using data taken by the Large Hadron Collider \cite{Pierog:17}. Although the differences 
between the shower observables predicted by different models have been reduced, there still remain 
some discrepancies (see Ref.~\cite{Pierog:17} for details).    

The mean value of $X_{max}$ obtained by Auger \cite{AugerXmax:14}, interpreted by using the updated 
versions of current hadronic interaction models, shows that the composition is light from $\sim 10^{18}$ 
up to $\sim 10^{18.6}$ eV. From $\sim 10^{18.3}$ eV, the composition becomes progressively heavier for 
increasing values of the primary energy. This trend is consistent with the results obtained by using the 
standard deviation of the $X_{max}$ distribution \cite{AugerXmax:14}. Therefore, if the shower predictions, 
based on the current high energy hadronic interaction models, are not too far from the correct description, 
it can be said that there is evidence of the existence of a light component that dominates the spectrum 
below the ankle. On the other hand, the $X_{max}$ parameter reconstructed from the data taken by the 
fluorescence telescopes of Telescope Array is also compatible with a light composition below the ankle, 
when it is interpreted by using the current hadronic interaction models \cite{TA:15}. It is worth mentioning 
that the $X_{max}$ distributions, as a function of primary energy, obtained by Auger and Telescope Array 
are compatible within systematic uncertainties \cite{Souza:17}. However, the presence of heavier primaries 
above the ankle cannot be confirmed by the Telescope Array data due to the limited statistics of the event 
sample \cite{Souza:17}.         

The Auger data show that the large scale distribution of the cosmic ray arrival directions is compatible 
with an isotropic flux, in the energy range from $\sim 10^{18}$ eV up to the ankle \cite{Auger:12}. This 
result is incompatible with a galactic origin of the light component that seems to dominate the flux in 
this energy range \cite{Auger:12}. Therefore, the scenarios in which the ankle is interpreted as the point 
in which the galactic to extragalactic transition takes place are incompatible with present data, assuming 
that the $X_{max}$ predictions based on current high energy hadronic interaction models are close to the 
real ones. 

There are two main scenarios that can explain the experimental data. In the first one, the light component 
below the ankle corresponds to a different population of sources than the ones that are responsible of the 
flux above the ankle \cite{Aloisio:14}. In this model, the spectral index of the spectrum injected by the 
sources that dominate the flux below the ankle is steeper than the one corresponding to the other population. 
In the second scenario, the light component originates from the photodisintegration of high energy and heavier 
nuclei in a photon field present in the environment of the source. This has been proposed as a general 
mechanism \cite{Unger:15} that can take place in starburst galaxies \cite{Anchordoqui:17} and also in the 
context of the UHECR acceleration in $\gamma$-ray bursts \cite{Globus:15a,Globus:15b}. Also, in 
Ref.~\cite{Kachel:17} a model combining photodisintegration and hadronic interactions of UHECRs in the 
photon and proton gases present in the central regions of active galaxies has been proposed.    
 
In this work, we study the possibility of the formation of the extragalactic light component that seems to 
dominate the UHECR flux below the ankle by the photodisintegration of heavier and more energetic nuclei 
in the radiation field present in the central region of active galaxies. In this scenario, the UHECRs are 
accelerated near the supermassive black hole present in this type of galaxy. In this work, the propagation 
in the source environment is modeled in more detail than in Ref.~\cite{Kachel:17}, including a 
three-dimensional simulation of the propagation of the UHECR nuclei in the random magnetic field and the
photon gas present in the source environment. Also, the conditions by which these types of models can 
properly describe the present experimental results are discussed.

\section{Cosmic ray propagation in the source environment}

As mentioned before, the case in which the UHECRs are accelerated in the central 
region of an active galaxy is considered. Therefore, after escaping from the 
acceleration zone, the cosmic rays propagate through a region that is filled with 
a low energy photon gas and a turbulent magnetic field. 

The simulation of the propagation of the cosmic ray nuclei in the source 
environment is performed by using a dedicated program specifically 
developed for that purpose. The interactions with the low energy photon 
gas implemented are photodisintegration and photopion production. This
implementation is based on the CRPropa 3 program \cite{CRPropa3:16}. The 
photopion production is performed by using the SOPHIA code \cite{sophia}.
Nuclear and neutron decay are also included in the simulation. The 
propagation of the particles is three dimensional, which includes the 
deflection of the charged particles in the turbulent magnetic field (see 
below).      

The source environment is modeled as a spherical region of radius $R=10^{-7}$ kpc 
$\cong 3.1\times10^{14}$ cm \cite{Kachel:09}. The numerical density of the low 
energy photon gas is taken as a broken power law \cite{Szabo:94,Unger:15},
\begin{equation}
\frac{dn}{d \varepsilon}(\varepsilon) = n_b \left\{
\begin{array}{ll}
\left( \mathop{\displaystyle \frac{\varepsilon}{\varepsilon_b} } \right)^{\alpha} & \varepsilon\leq \varepsilon_b \\[0.4cm]
\left( \mathop{\displaystyle \frac{\varepsilon}{\varepsilon_b} } \right)^{\beta} & \varepsilon_b < \varepsilon_b \leq \varepsilon_{max} \\[0.4cm]
0 & \varepsilon > \varepsilon_{max}
\end{array}
\right.,
\label{dnde}
\end{equation}  
where the spectral indexes are taken from Ref.~\cite{Szabo:94}, $\alpha=3/2$ and $\beta=-2$, 
and the energy break and the maximum energy are determined following Ref.~\cite{Kachel:09}, 
$\varepsilon_b=0.2$ eV and $\varepsilon_{max}=5$ eV.

It is worth mentioning that the luminosity of the low energy photon gas is related to the 
normalization $n_b$ through the following expression,
\begin{equation}
L=c\, \pi R^2 \varepsilon_b^2 \left[ \frac{2}{7}+\ln\left( \frac{\varepsilon_{max}}{\varepsilon_b} \right) \right] n_b,
\end{equation}    
where $c$ is the speed of light.

The interaction length $\lambda_I$ of a nucleus propagating in a photon gas is given by
\begin{equation}
\frac{1}{\lambda_{I}(E)} = \frac{1}{2\, \gamma^2} \int_0^\infty d\varepsilon \, \frac{dn}{d\varepsilon}(\varepsilon)\, \varepsilon^{-2}%
\int_0^{2\, \gamma \varepsilon} d\varepsilon' \varepsilon' \sigma(\varepsilon'), 
\end{equation}
where $\gamma$ is the Lorentz factor of the nucleus and $\sigma(\varepsilon')$ is the photo-nuclear 
interaction cross section for a photon of energy $\varepsilon'$ in the rest frame of the nucleus.
The interaction lengths corresponding to the photon gas density of Eq.~(\ref{dnde}) are calculated 
by using the tools developed for CRPropa 3, which are accessible at \cite{CRPropaTab}.

The propagation of the charged nuclei in the random magnetic field is performed following the method 
developed in Ref.~\cite{Protheroe:02}. The propagation is described by a three-dimensional random 
walk in which the directions of the particles change according to the scattering length $\lambda_{SL}=3D/c$, 
where $D$ is the spatial diffusion coefficient. The distance traveled by the particles after being
scattered by the magnetic field is sampled from an exponential distribution with the mean value given 
by $\bar{\ell}=\bar{\theta}^2 \lambda_{SL}$, where $\bar{\theta}$ is the mean value of the exponential 
distribution from which the angular change in the direction of propagation is sampled. In 
Ref.~\cite{Protheroe:02}, it is found that the method renders accurate results for 
$\bar{\theta} < 0.09$ rad ($5^{\circ}$). 

The diffusion coefficient used in the simulations is taken from Ref.~\cite{Harari:14}, 
which is given by
\begin{equation}
D(E)=\frac{c}{3} l_c \left[4 \left( \frac{E}{E_c} \right)^2 +a_I \left( \frac{E}{E_c} \right) + a_L \left( \frac{E}{E_c} \right)^{2-m}   \right],
\end{equation}       
where $l_c$ is the coherent length of the random magnetic field. Here $E_c=Z e B\, l_c$, where 
$Z$ is the charge number of the nucleus, $e$ is the absolute value of the electron charge, and 
$B=\sqrt{\langle B^2(x) \rangle}$ is the root mean square of the random magnetic field. The 
parameter $m$ and the numerical values of the parameter $a_I$ and $a_L$ depend on the type of
turbulence, and for a Kolmogorov spectrum, which is the one considered in this work, these three
parameters take the following values \cite{Harari:14}: $m=5/3$, $a_L=0.23$, and $a_I=0.9$.    

The propagation of charged particles in a random magnetic field depends on the  distance
of the particles under the influence of the field. For traveled distances much smaller than the 
scattering length $\lambda_{SL}$, the propagation is ballistic, and for traveled distances much 
larger than $\lambda_{SL}$, the propagation is diffusive (see, for instance, Ref.~\cite{Harari:14}). 
It is worth mentioning that by using the method for the propagation of charged particles in a 
random magnetic field developed in Ref.~\cite{Protheroe:02}, the two different regimes of propagation 
are included. 

The simulation starts by injecting a nucleus of certain type and energy at the center of a sphere 
of radius $R$ and ends when all particles leave the sphere. The propagation of a particle proceeds 
as follows. Let us consider a particle in a given position $\vec{x}$ inside the sphere with a 
given velocity $\vec{v}=c\, \hat{n}$, where $\hat{n}$ is a unit vector (it is assumed that all particles 
move at the speed of light). In the next step, the position of the particle is 
$\vec{x}'=\vec{x}+\Delta s\, \vec{n}$, where $\Delta s$ is obtained by sampling the exponential 
distribution with mean $\bar{\ell}=\bar{\theta}^2 \lambda_{SL}$ for which $\bar{\theta}$ is chosen in 
such a way that it fulfils two conditions: it is smaller than 0.09 rad (see above) and also it is small 
enough in such way that $\Delta s \ll \lambda_{T}$, where 
$\lambda_{T}^{-1}=\lambda_{PD}^{-1}+\lambda_{PP}^{-1}+\lambda_{D}^{-1}$. Here $\lambda_{PD}$ is the 
photodisintegration interaction length, $\lambda_{PP}$ is the photopion production interaction length, 
and $\lambda_{D}$ is the decay length of the nucleus. The particle at position $\vec{x}'$ can interact, 
decay, or change its direction of motion. In order to decide the outcome, an integer number is taken at 
random from a Poisson distribution with mean value $\mu=\Delta s /\lambda_T$. If this number is one, the 
particle interacts or decays, if not its direction of motion is modified in such a way that the new 
velocity vector forms an angle $\theta$ with the velocity vector at position $\vec{x}$. The $\theta$ angle 
is obtained by sampling an exponential distribution with mean value $\bar{\theta}$ (see above). In the case 
that the particle decays or interacts, three distances are sampled from three different exponential distributions 
with mean values $\lambda_{PD}, \lambda_{PP}$, and $\lambda_{D}$, and thus the process undergone by the particle 
is the one corresponding to the smallest distance. As mentioned before, the implementation of the 
photodisintegration, photopion production, and nuclear and neutron decay are based on the CRPropa 3 program.         

\begin{figure}[!ht]
\includegraphics[width=8cm]{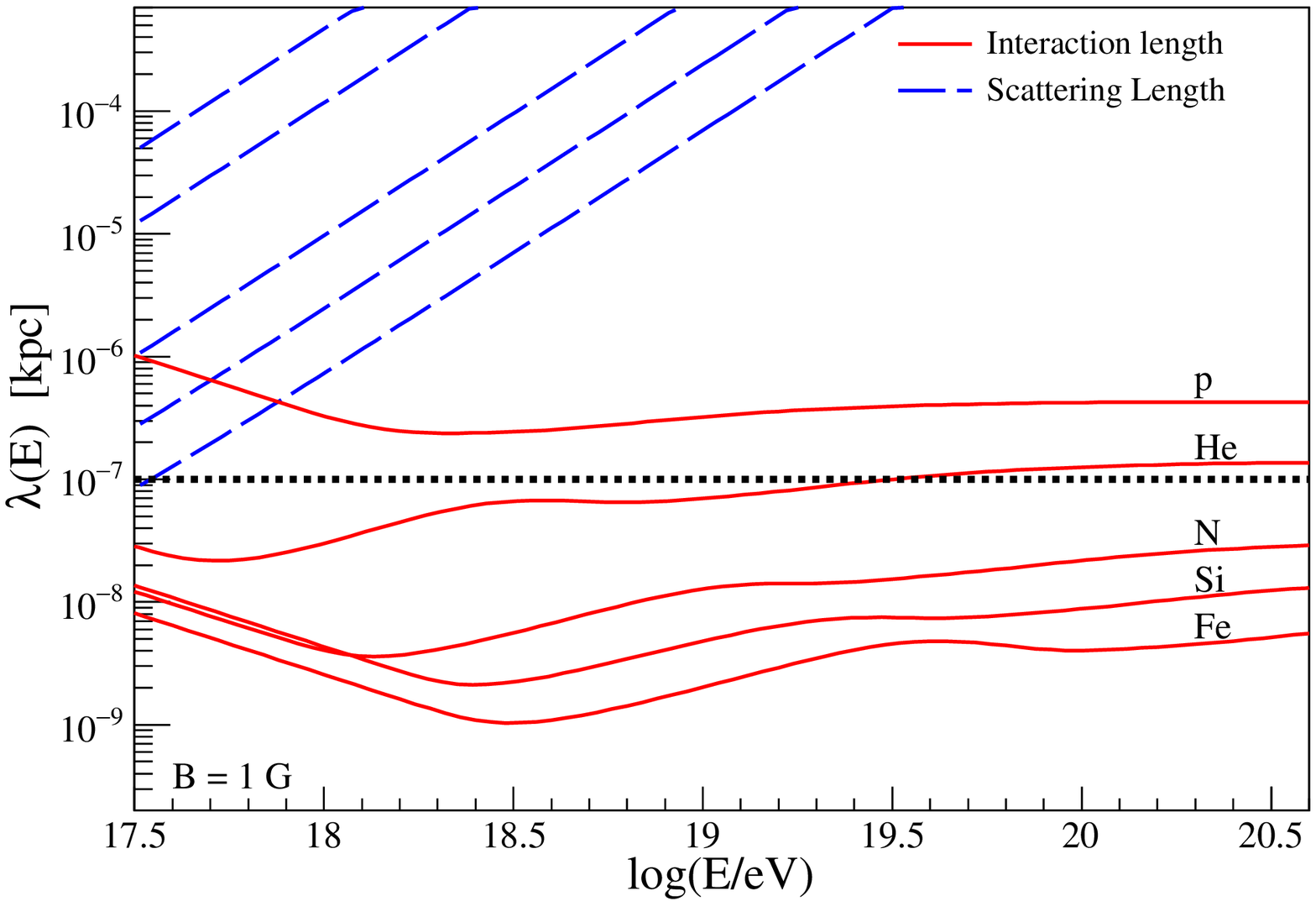}
\includegraphics[width=8cm]{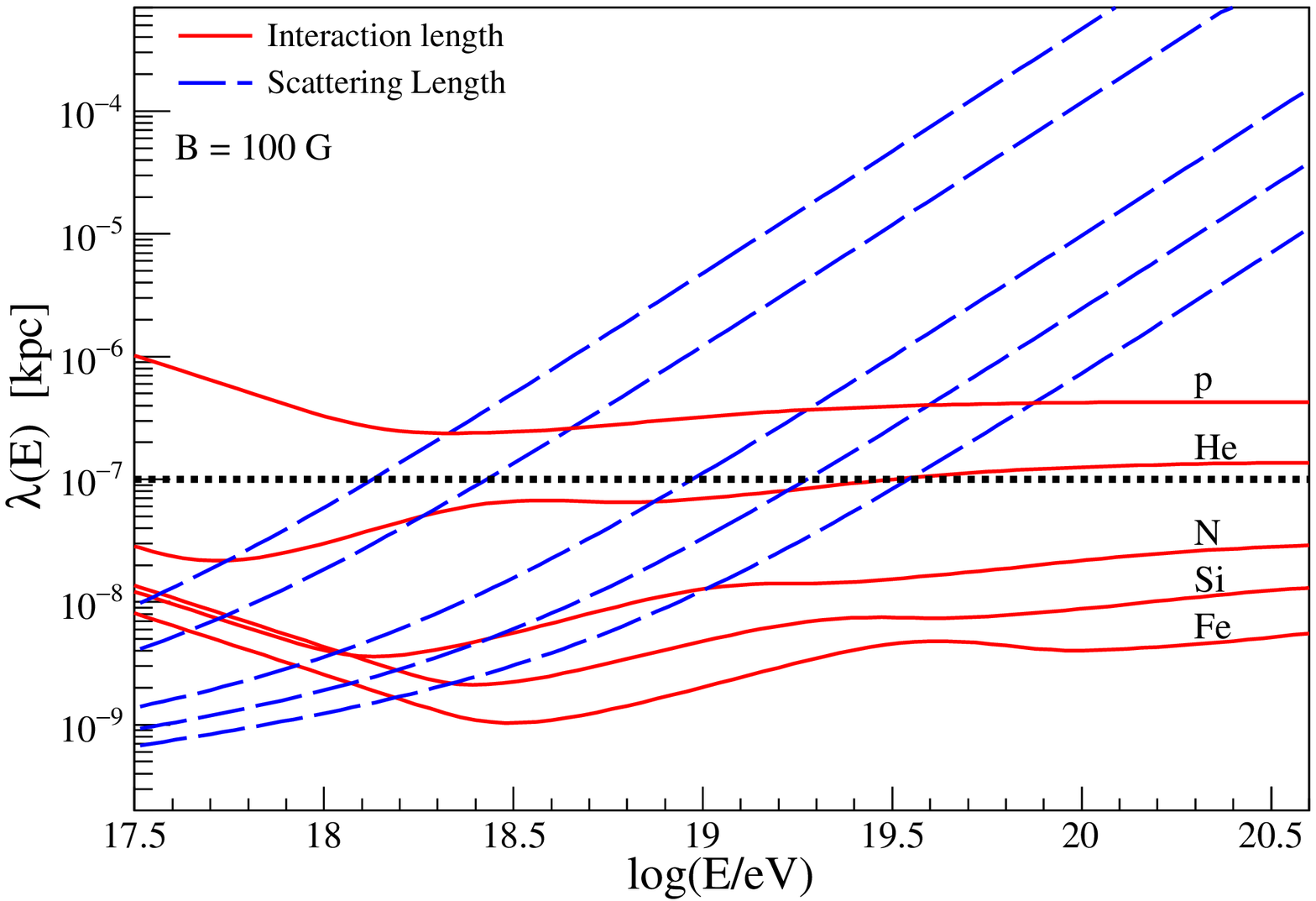}
\caption{Interaction and scattering lengths of proton (p), helium (He), nitrogen (N), silicon (Si), and 
iron (Fe) as a function of energy. In both cases, the curves are ordered from bottom to top by decreasing 
primary mass. The dotted line corresponds to the radius of the sphere. (Top) $B=1$ G. (Bottom) 
$B=100$ G. The coherence length of the random magnetic field is $l_c=R/10$. \label{SILengths}}
\end{figure}

Figure \ref{SILengths} shows the interaction lengths and the scattering lengths in the random magnetic 
field for five different types of nuclei: proton (p), helium (He), nitrogen (N), silicon (Si), and iron (Fe). 
The interaction length includes the photodisintegration and photopion production processes. Note that these 
curves present the ``L'' shape mentioned in Ref.~\cite{Unger:15}. The dotted lines in the plots correspond to 
the radius of the sphere, $R$. The top panel of the figure corresponds to a luminosity of 
$L=10^{41}$ erg s$^{-1}$, a random magnetic field of $B=1$ G, and coherence length $l_c=R/10$. In this case, 
the interaction length of protons is larger than the radius of the sphere in the energy range under consideration. 
Also, the interaction length of helium is larger than the radius of the sphere above $\sim 10^{19.5}$ eV. However, 
the interaction lengths of the other three species considered are smaller than the radius of the sphere in the 
whole energy range. The scattering lengths of all nuclear species considered are larger than $R$, and then the 
propagation of all nuclei is mainly ballistic in the energy range under consideration. Therefore, proton and 
high energy helium nuclei are less affected than the other nuclear species by photodisintegration and photopion 
production processes.          

The bottom panel of the figure shows the interaction and the scattering lengths, but for $B=100$ G and 
$l_c=R/10$. As can be seen from the plot, the propagation in the random magnetic field of all nuclear 
species considered is diffusive at low energies and ballistic at high energies. Therefore, in general, 
the nuclei stay inside the sphere more time than in the case corresponding to $B=1$ G, and then the 
composition of the nuclei that leave the sphere becomes lighter compared with the one corresponding 
to that case.       

It is assumed that the cosmic ray nuclei are accelerated in a region close to a supermassive black hole,
in such a way that the energy spectrum is given by a power law with an exponential cutoff,
\begin{equation}
\varphi(E) = \varphi_0\, E^{-\Gamma} \exp\left( -\frac{E}{Z\, E_{max}^{p}} \right),
\label{SpecInj}
\end{equation}  
where $\varphi_0$ is a normalization constant, $\Gamma$ is the spectral index, $E_{max}^{p}$ is the 
maximum energy for the proton component, and $Z$ is the charge number of the nucleus. Note that the 
cutoff energy is proportional to the charge number, which is motivated by acceleration processes of 
electromagnetic origin \cite{Allard:05,AugerFit:17}. The spectral index is taken as $\Gamma=1$, which 
is motivated by acceleration mechanisms taking place in the accretion disks around massive black holes 
\cite{Blandford:76} and also by the fit of the flux and mass composition data obtained by Auger and 
reported in Ref.~\cite{AugerFit:17}. 

\begin{figure}[!ht]
\includegraphics[width=8cm]{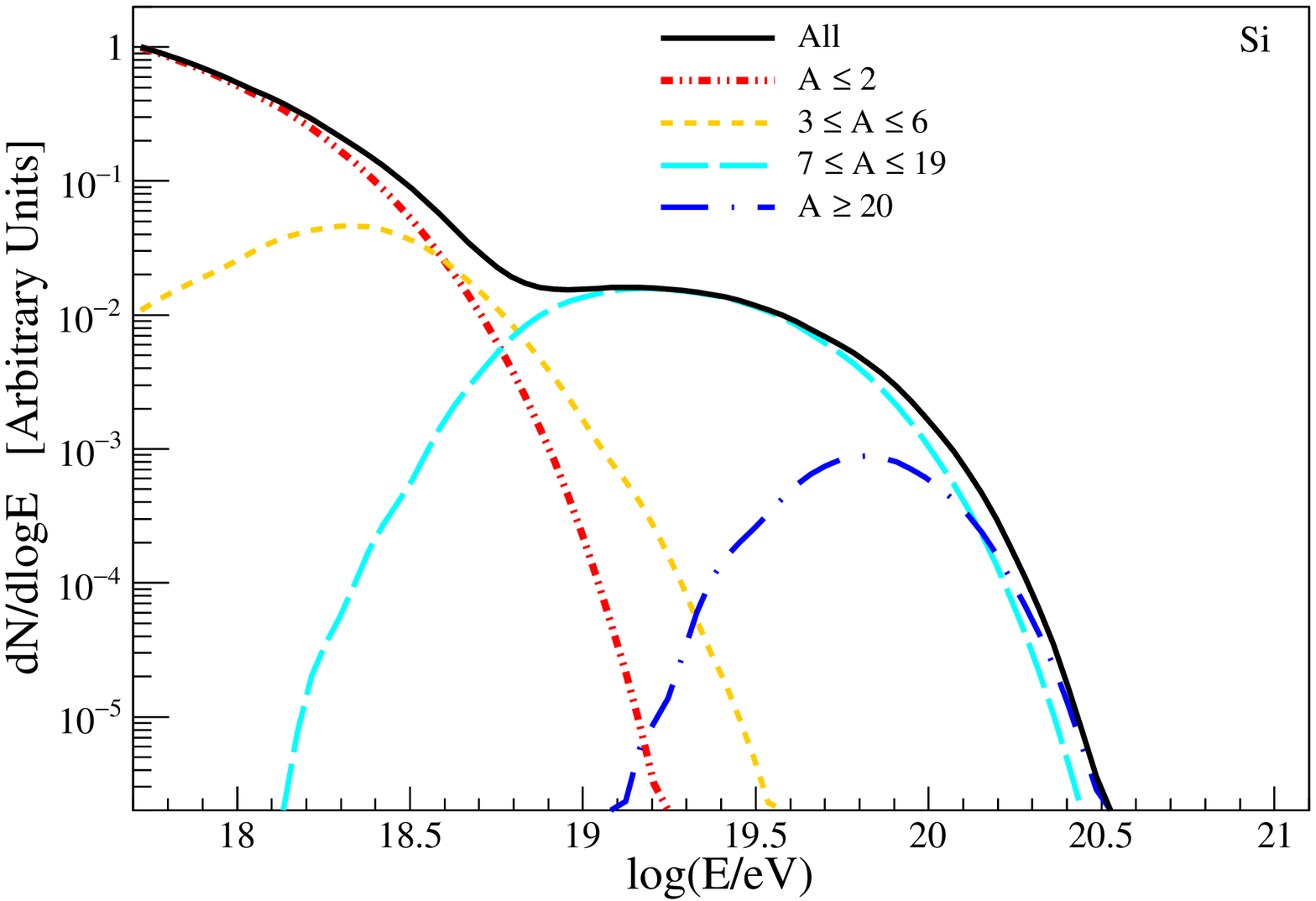}
\includegraphics[width=8cm]{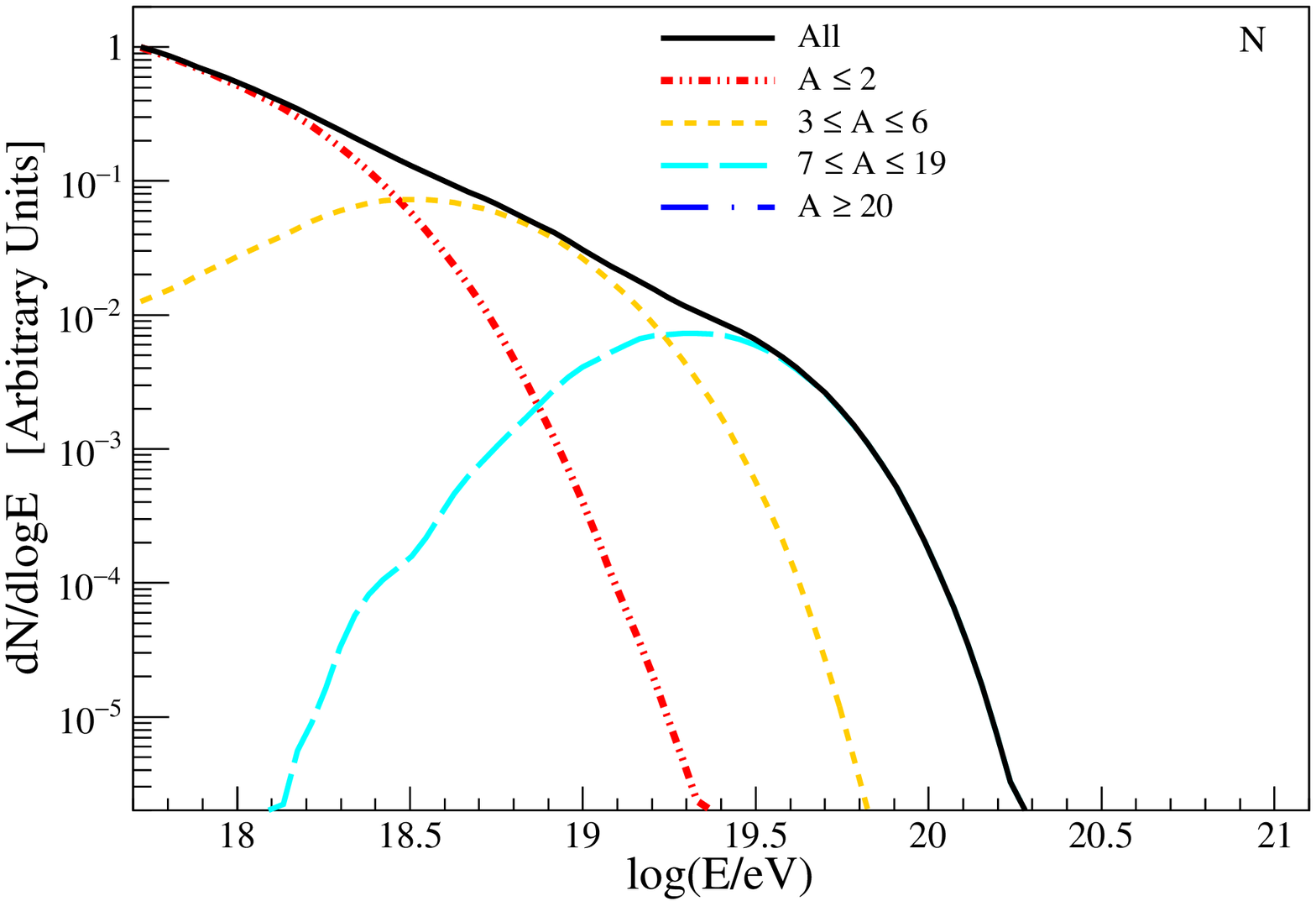}
\caption{Energy spectra of nuclei that leave the sphere corresponding to silicon (top) and nitrogen
(bottom). The magnetic field is such that $B=100$ G and $l_c=R/10$. The parameters of the injection 
spectrum are: $\Gamma=1$ and $E_{max}^{p}=10^{18.5}$ eV. \label{SpecSource}}
\end{figure}

Figure \ref{SpecSource} shows the energy spectra of the cosmic rays that leave the sphere corresponding 
to the injection of silicon (top panel) and nitrogen (bottom panel) nuclei at the center of the sphere. 
The injection spectrum of the nuclei is the one corresponding to Eq.~(\ref{SpecInj}) with 
$E_{max}^{p}=10^{18.5}$ eV. The magnetic field is such that $B=100$ G and $l_c=R/10$ (see bottom panel 
of Fig.~\ref{SILengths}).      
From the figure it can be seen that, for both nuclear species, a low energy light component is generated
due to the interactions undergone by the primary nuclei during propagation through the sphere.


\section{Flux at Earth}

The cosmic rays that leave the source environment are injected in the intergalactic medium and propagated 
from a given position in the Universe to Earth. The propagation of the particles is performed by using 
CRPropa 3. The simulations include photopion production and photodisintegration in the cosmic microwave 
background (CMB) and in the extragalactic background light (EBL), pair production on the CMB and on the EBL, 
nuclear decay, and the effects of the expansion of the Universe. The intergalactic magnetic field intensity 
is assumed to be negligible and then the propagation is unidimensional. A uniform distribution of sources 
in the Universe is assumed and the EBL model used in the simulations is the one developed in 
Ref.~\cite{Kneiske:04}. The redshift range considered in the simulations starts at $z=0$ and ends at $z=5$. 

The production of UHECRs over cosmological timescales is unknown. This source evolution is accounted by 
a function of the redshift $z$, $S(z)$. In this work, two cases are considered. $S(z)=1$, which corresponds to
the case of no evolution of the sources and 
\begin{equation}
S(z) = \left\{
\begin{array}{ll}
\left( 1+z \right)^5 & z \leq 1.7 \\[0.2cm]
2.7^{5} & 1.7 < z \leq 2.7 \\[0.2cm]
2.7^{5} \times 10^{2.7-z} & z > 2.7
\end{array}
\right.,
\label{Sz}
\end{equation}  
which corresponds to the case of active galactic nuclei (AGN) of Ref.~\cite{Aloisio:15}.

In order to fit the cosmic ray energy spectrum above $E=10^{17.5}$ eV, a galactic low energy iron 
component \cite{Unger:15} is assumed. The flux at Earth is supposed to be a power law with an 
exponential cutoff \cite{Aloisio:14}, which is given by
\begin{equation}
J_{G}(E) = c_G\ E^{-\Gamma_G} \exp\left(-\frac{E}{E_{cut}} \right), 
\end{equation}
where $\Gamma_G = 3.29$ is the spectral index for energies below the ankle \cite{AugerICRC:17},  
$E_{cut}$ is the cutoff energy of the galactic component, and $c_G$ is a normalization constant. 
The last two parameters are chosen in each model considered in order to fit the Auger spectrum.

Following Ref.~\cite{AugerFit:17}, five nuclear species that are accelerated in the sources are 
considered: proton, helium, nitrogen, silicon, and iron.

The Auger energy spectrum reported in Ref.~\cite{AugerICRC:17} is fitted by minimizing the $\chi^2$
given by
\begin{equation}
\chi^2=\sum_{i=1}^N \frac{\left( j_i-J_G(c_G,E_i)-\sum_A c_A J_A(E_i) \right)^2}{\sigma_i^2},
\end{equation}
where $j_i$ and $\sigma_i$ are the measured flux and its uncertainty, respectively, corresponding to 
the energy bin centered at energy $E_i$. Here $N$ is the number of energy bins considered in the fit 
and $A=\{\textrm{p, He, N, Si, Fe}\}$. The free fit parameters are $c_G$ and $c_A$, i.e., just the 
relative contributions of the different components are fitted. Note that the fitting parameters have 
to be positive or zero. This condition is fulfilled during the minimization procedure.

\begin{figure}[!ht]
\includegraphics[width=8cm]{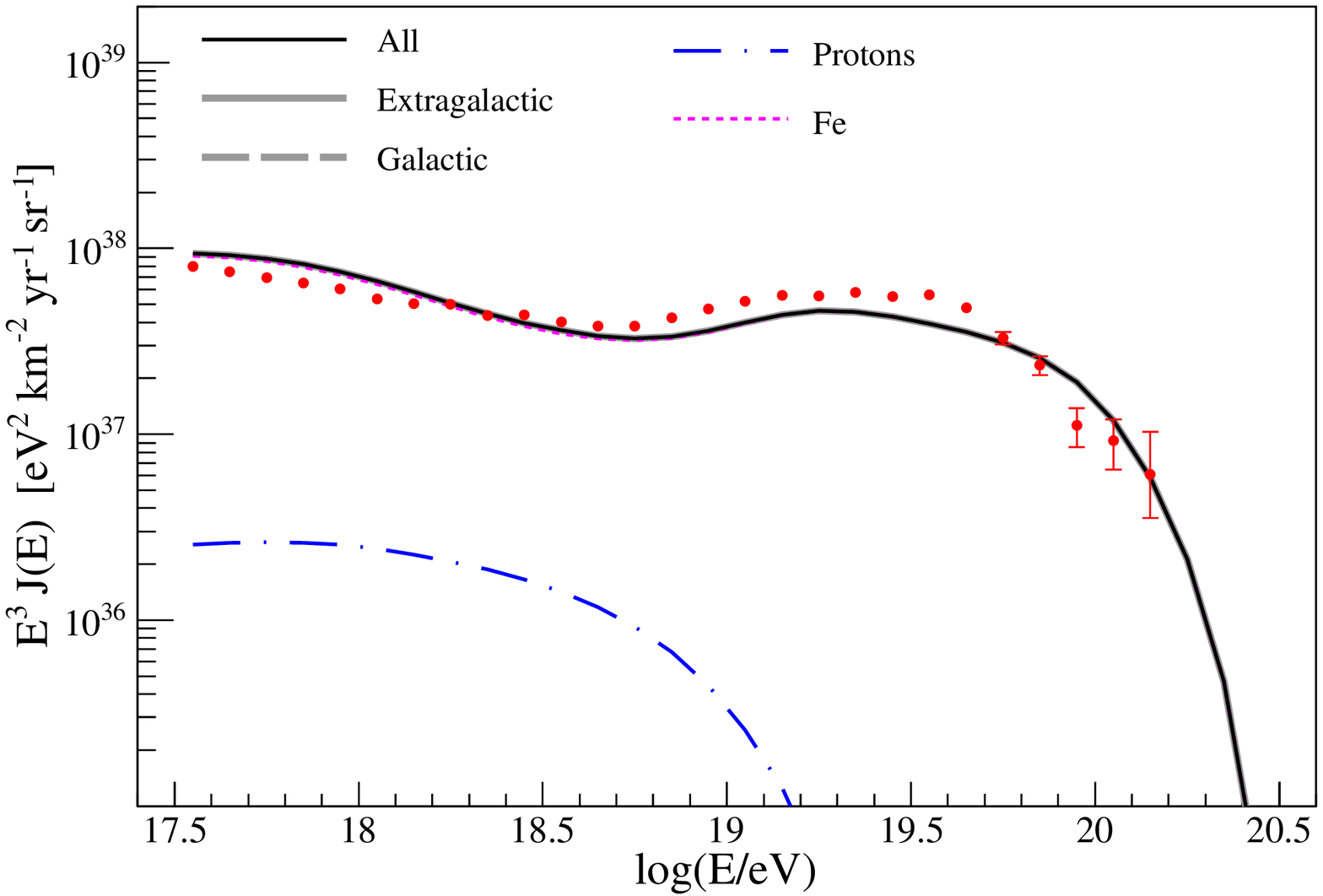}
\includegraphics[width=8cm]{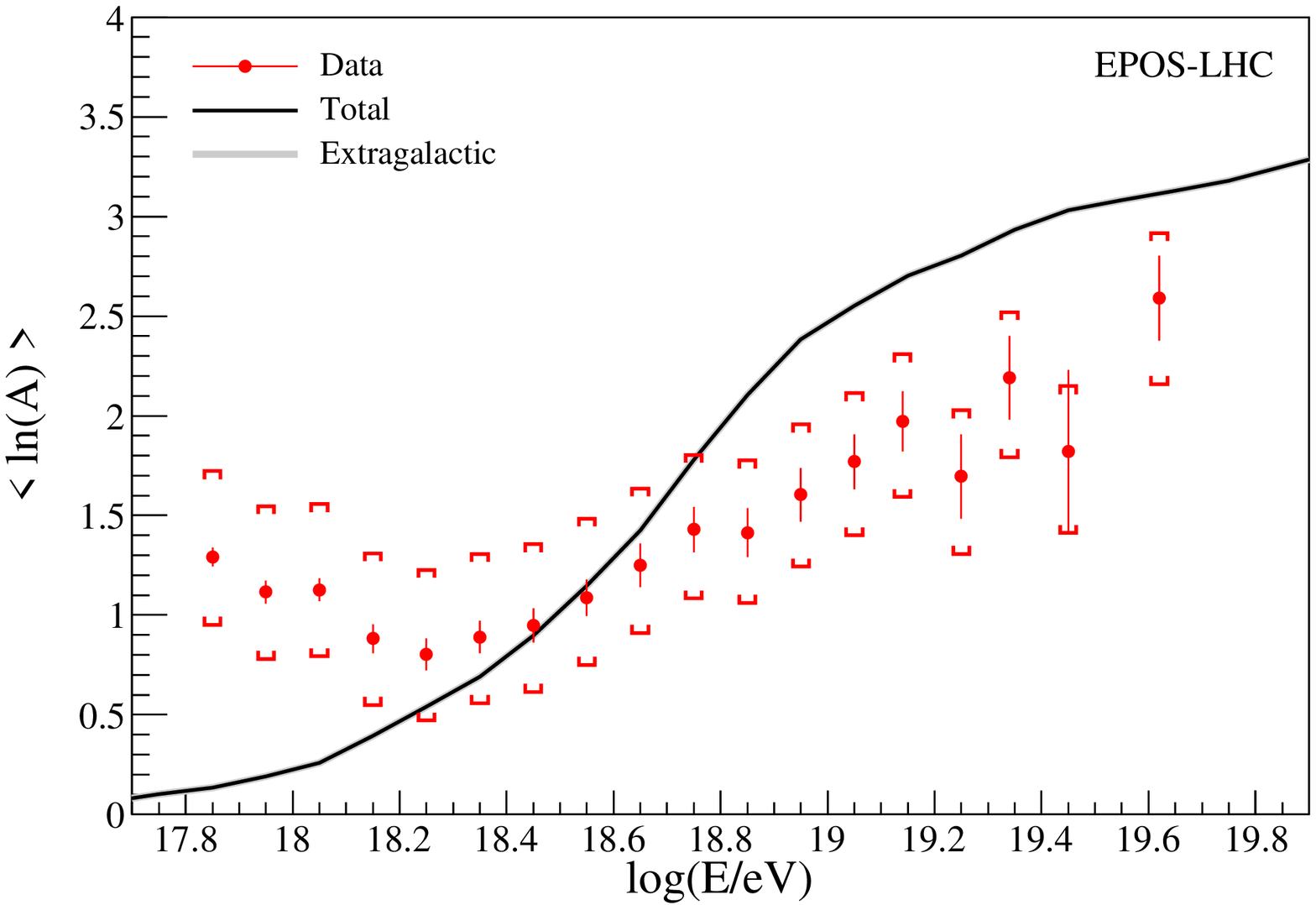}
\includegraphics[width=8cm]{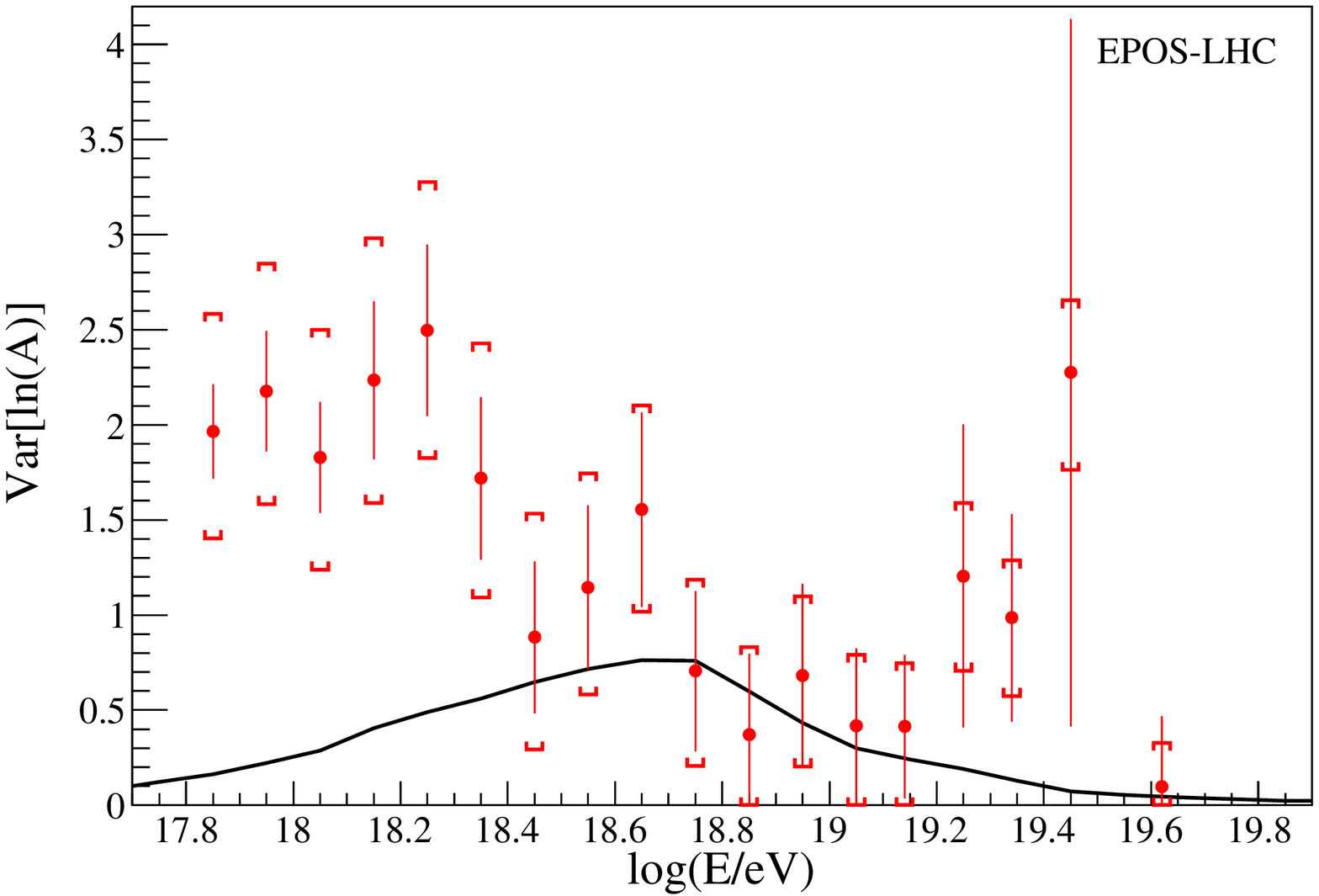}
\caption{Fit of the UHECR spectrum (top) and the prediction for the mean value of the natural logarithm 
of the mass number (middle) and its variance (bottom). The experimental data were obtained by 
Auger \cite{AugerICRC:17,AugerXmax:14} and the high energy hadronic interaction model used in the composition 
analysis is EPOS-LHC. The parameters of the model are $L=10^{41}$ erg s$^{-1}$, $B=100$ G, $l_c=R/10$, 
$E_{max}^{p}=10^{18.5}$ eV, and $S(z)$ from Eq.~(\ref{Sz}).  \label{M1}}
\end{figure}

Figure \ref{M1} shows the fit of the Auger spectrum and the predicted mean value of the natural logarithm 
of the mass number $\langle \ln(A) \rangle$ and its variance $\textrm{Var}[\ln(A)]$ as a function of 
the primary energy compared with the experimental data obtained by Auger. The data points corresponding 
to the mean value of the natural logarithm of the mass number and its variance are obtained from the 
mean value and the variance of the $X_{max}$ parameter, which is reconstructed in an event-by-event 
basis from the data taken by the fluorescence telescopes of Auger \cite{AugerXmax:14}. Both quantities 
are obtained in Ref.~\cite{AugerXmax:14} by using simulations of the showers with the high energy 
hadronic interaction model EPOS-LHC \cite{EPOSLHC:15}. It is worth mentioning that, 
$\langle \ln(A) \rangle$ as a function of primary energy, obtained by using EPOS-LHC, falls in between 
the ones corresponding to the two other models more frequently used in the literature, QGSJET-II-04 
\cite{QGII:11,QGII:11b} and  Sibyll 2.3c \cite{Sibyll2.3c:15}. Moreover, $\langle \ln(A) \rangle$ 
obtained by using Sibyll 2.3c is above the one corresponding to EPOS-LHC, which in turn is above the 
one corresponding to QGSJET-II-04.  

The model of Fig.~\ref{M1} assumes a luminosity of the photon gas $L=10^{41}$ erg s$^{-1}$, a random 
magnetic field of the source environment such that $B=100$ G and $l_c=R/10$, and the maximum energy of the 
injected proton component $E_{max}^{p}=10^{18.5}$ eV, i.e., the parameters used to obtain Fig.~\ref{SpecSource}. 
The source evolution function considered is the one in Eq.~(\ref{Sz}). In the best fit scenario, the injected 
composition is dominated by iron nuclei with a small contribution of protons, as can be seen from the top panel 
of the figure, in which the contribution of the two components, obtained after propagation in the source 
environment and in the intergalactic medium, are shown. Note that, in this scenario, the galactic component 
appears at energies below $10^{17.5}$ eV. As can be seen from the figure, this model is not compatible with the 
Auger data. The reason for that is the very fast evolution of the sources at low redshift values, which increases 
the light component below the ankle, making the flux steeper than the one observed. Also the composition becomes 
progressively light for decreasing values of primary energy, which is inconsistent with the minimum in the 
$\langle \ln(A) \rangle$ observed at $10^{18.25}$ eV.      

\begin{figure}[!ht]
\includegraphics[width=8cm]{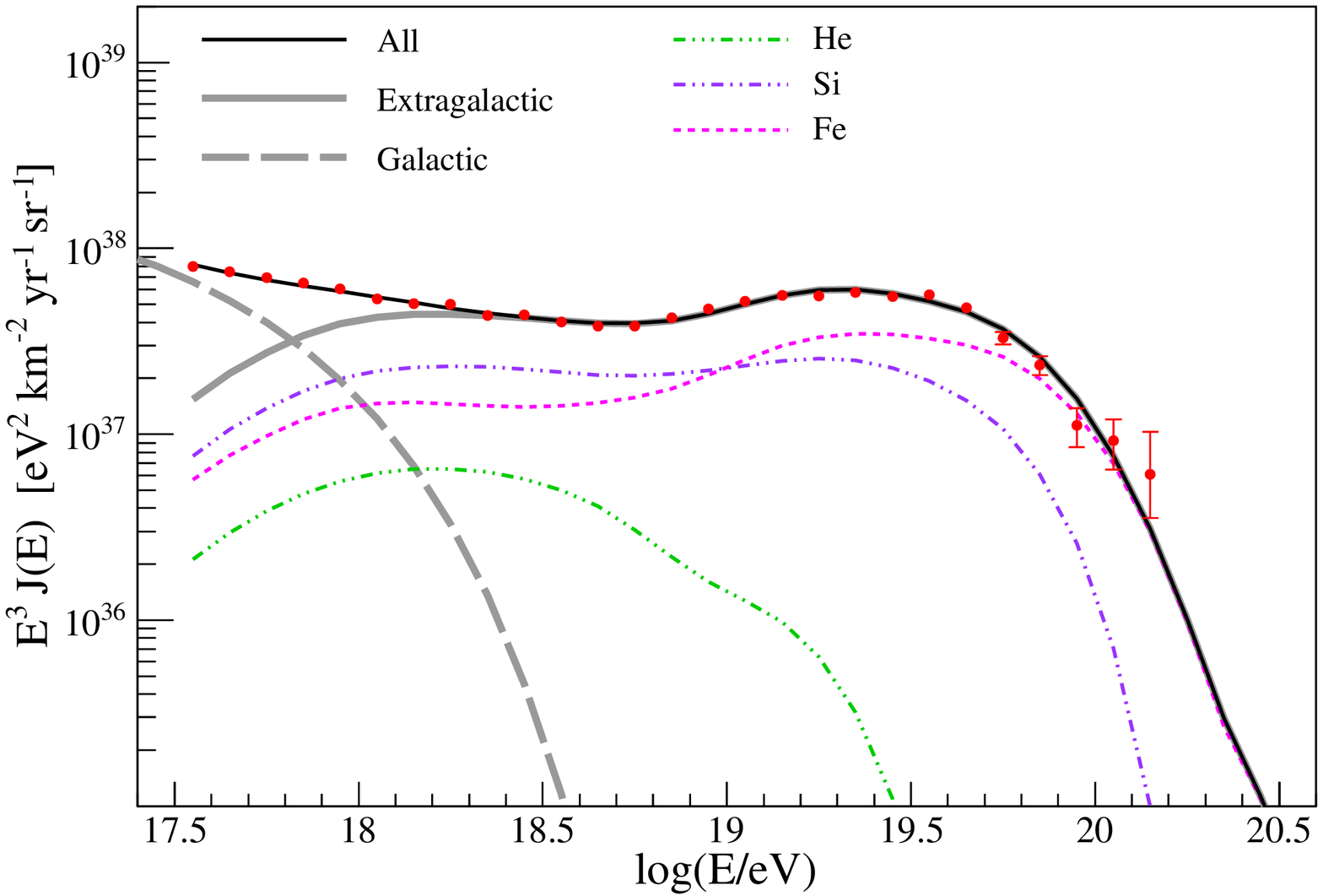}
\includegraphics[width=8cm]{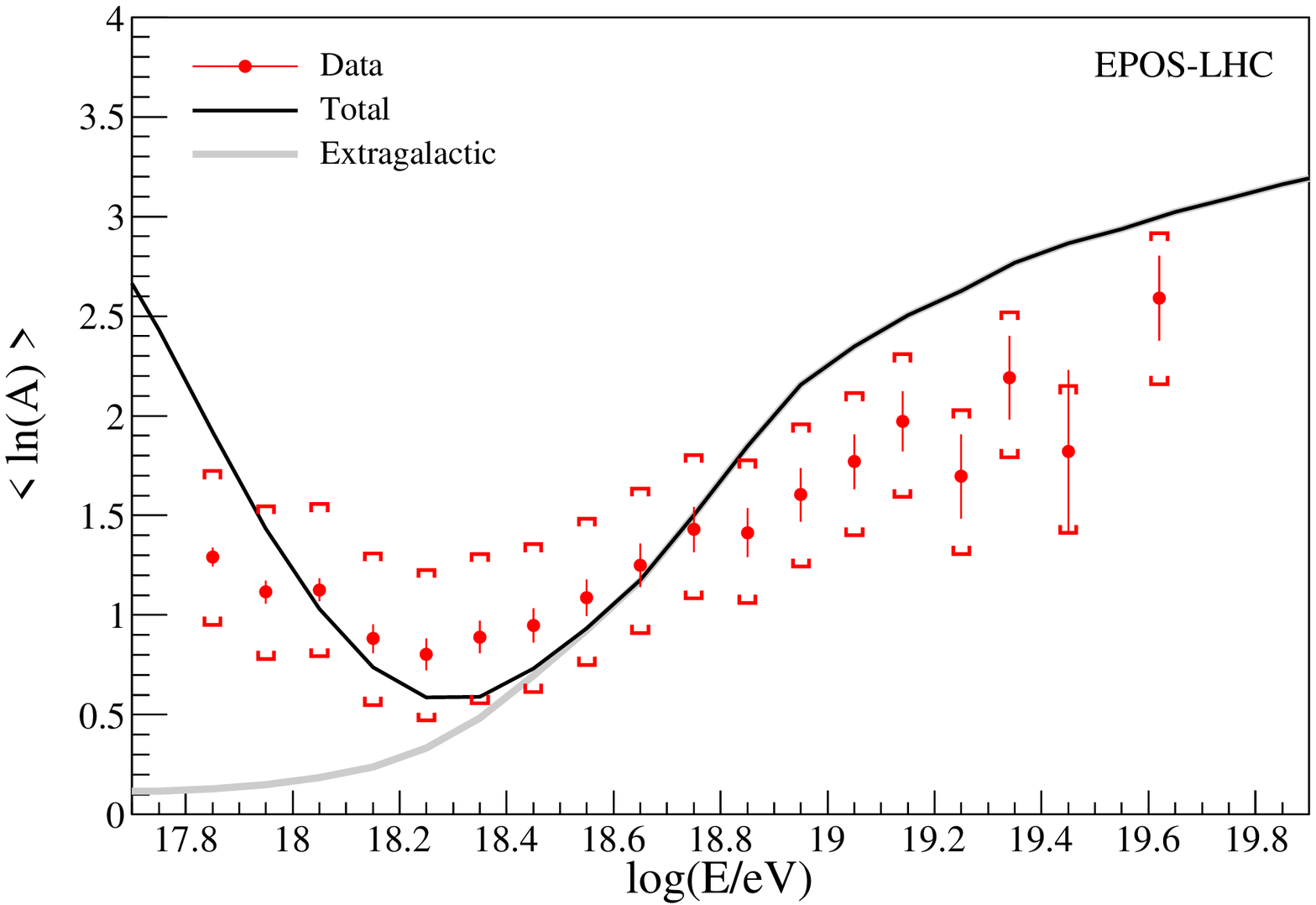}
\includegraphics[width=8cm]{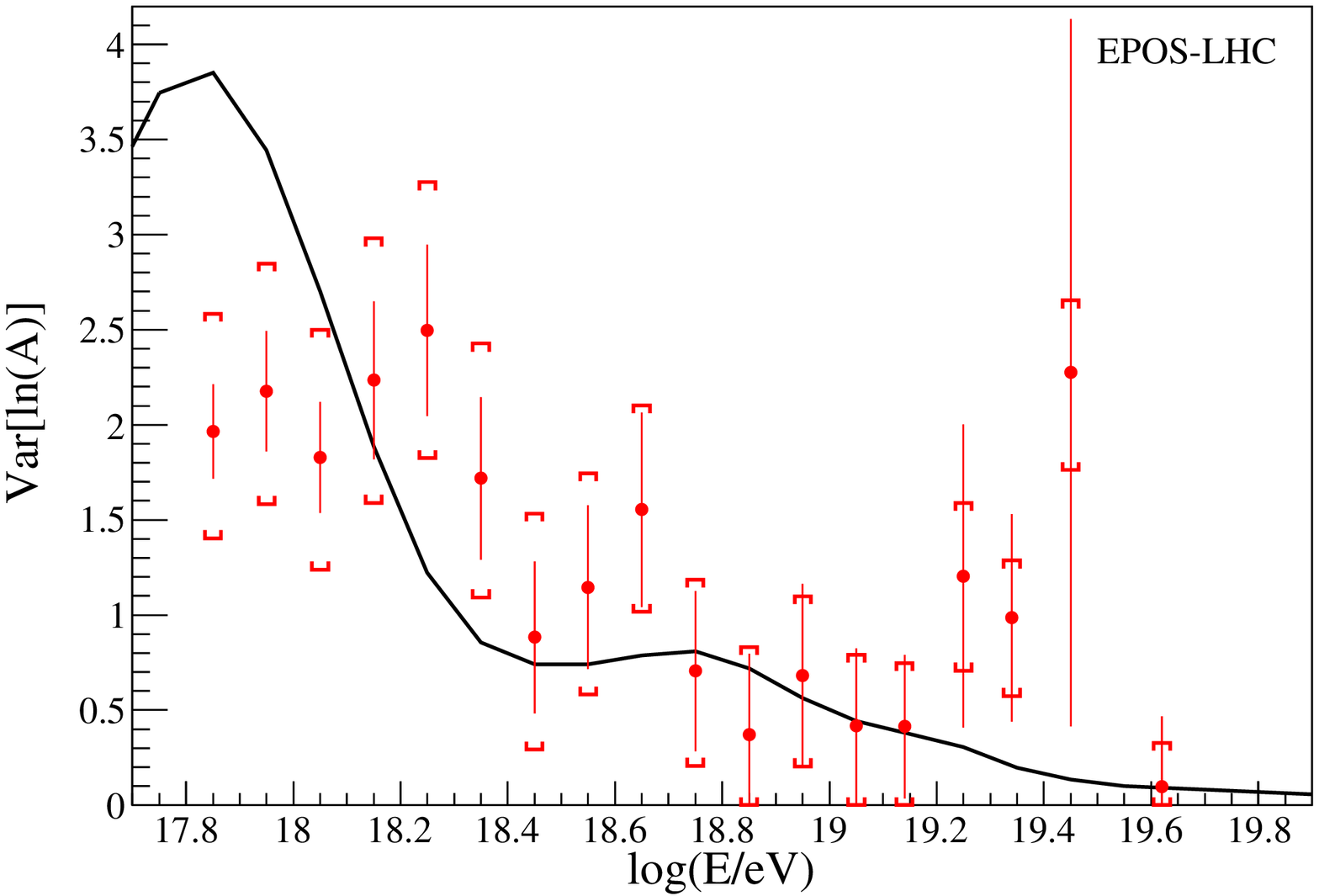}
\caption{Same as Fig.~\ref{M1}, but for $B=1$ G and $S(z)=1$. \label{M2}}
\end{figure}

Considering the same scenario as before but for the case in which the sources do not evolve with redshift, 
i.e.~$S(z)=1$, and for $B=1$ G, a good fit of the energy spectrum is obtained. In this case, the propagation in 
the source environment is not affected by the random magnetic field, as can be seen from the top panel of 
Fig.~\ref{SILengths}. The results corresponding to this model are shown in Fig.~\ref{M2}. In this case, the 
injection spectrum is dominated by helium, silicon, and iron; the proton and nitrogen contributions are 
negligible. In the top panel of the figure, the contributions of these three different nuclear species are 
shown, which are obtained after propagation in the source environment and in the intergalactic medium. In 
this case, the galactic component is not negligible in the energy range considered, as can be seen from the 
top panel of the figure (dashed line), and is such that $E_{cut}=10^{17.75}$ eV. Note that the discrepancies 
between the model predictions for $\textrm{Var}[\ln(A)]$ and the experimental data are larger at low energies, 
and this can be due to the too simple assumption for the composition of the galactic flux.     

\begin{figure}[!ht]
\includegraphics[width=8cm]{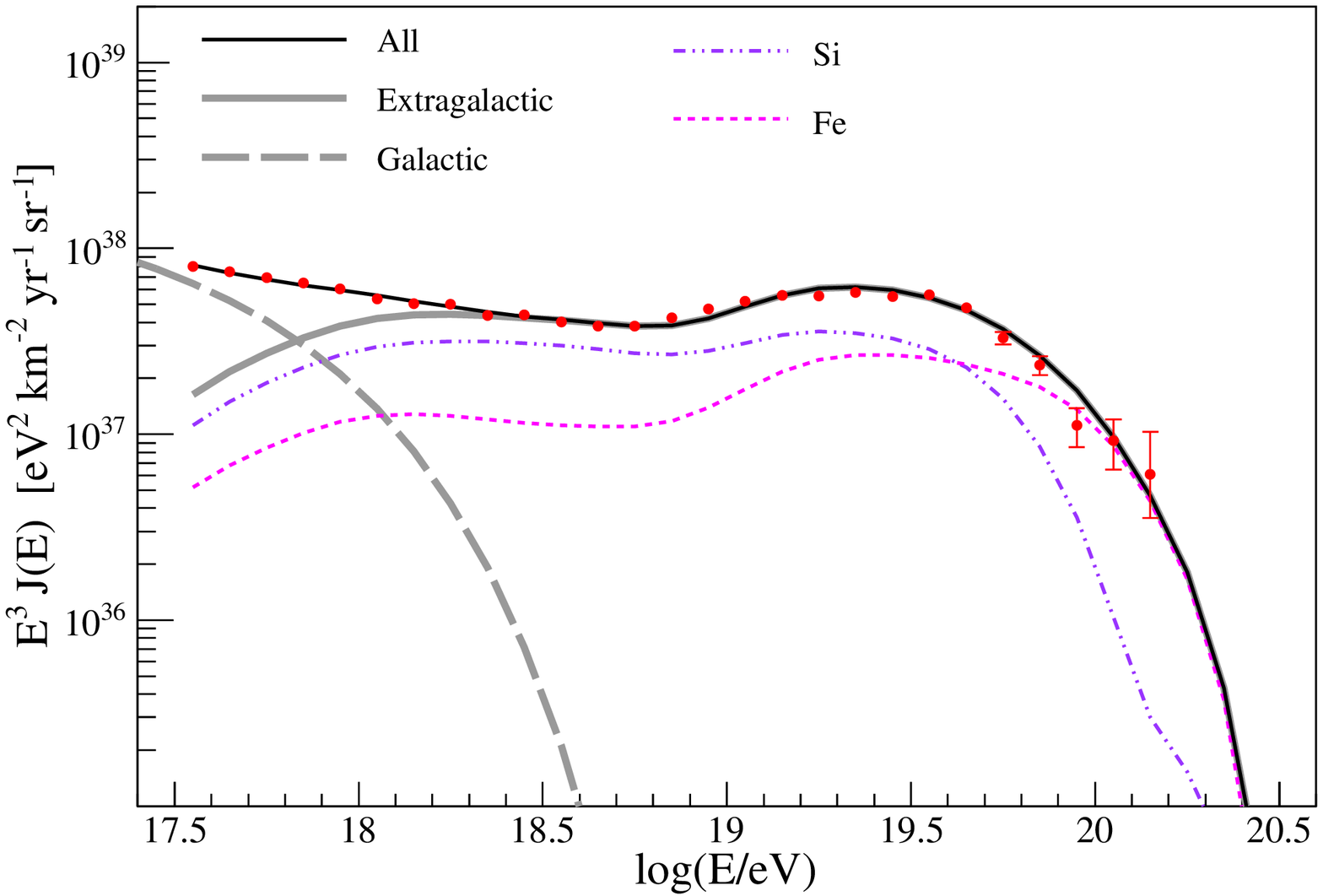}
\includegraphics[width=8cm]{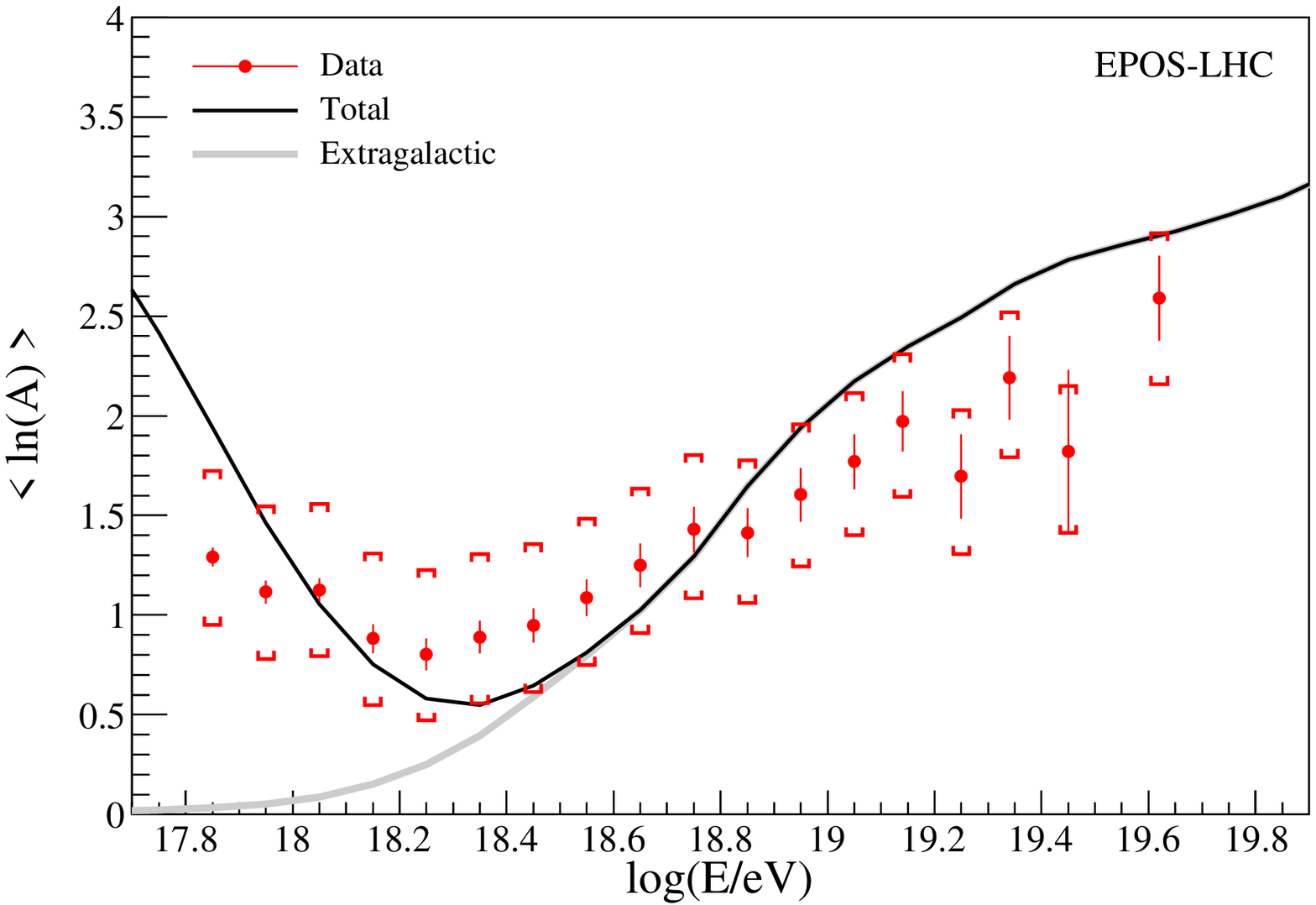}
\includegraphics[width=8cm]{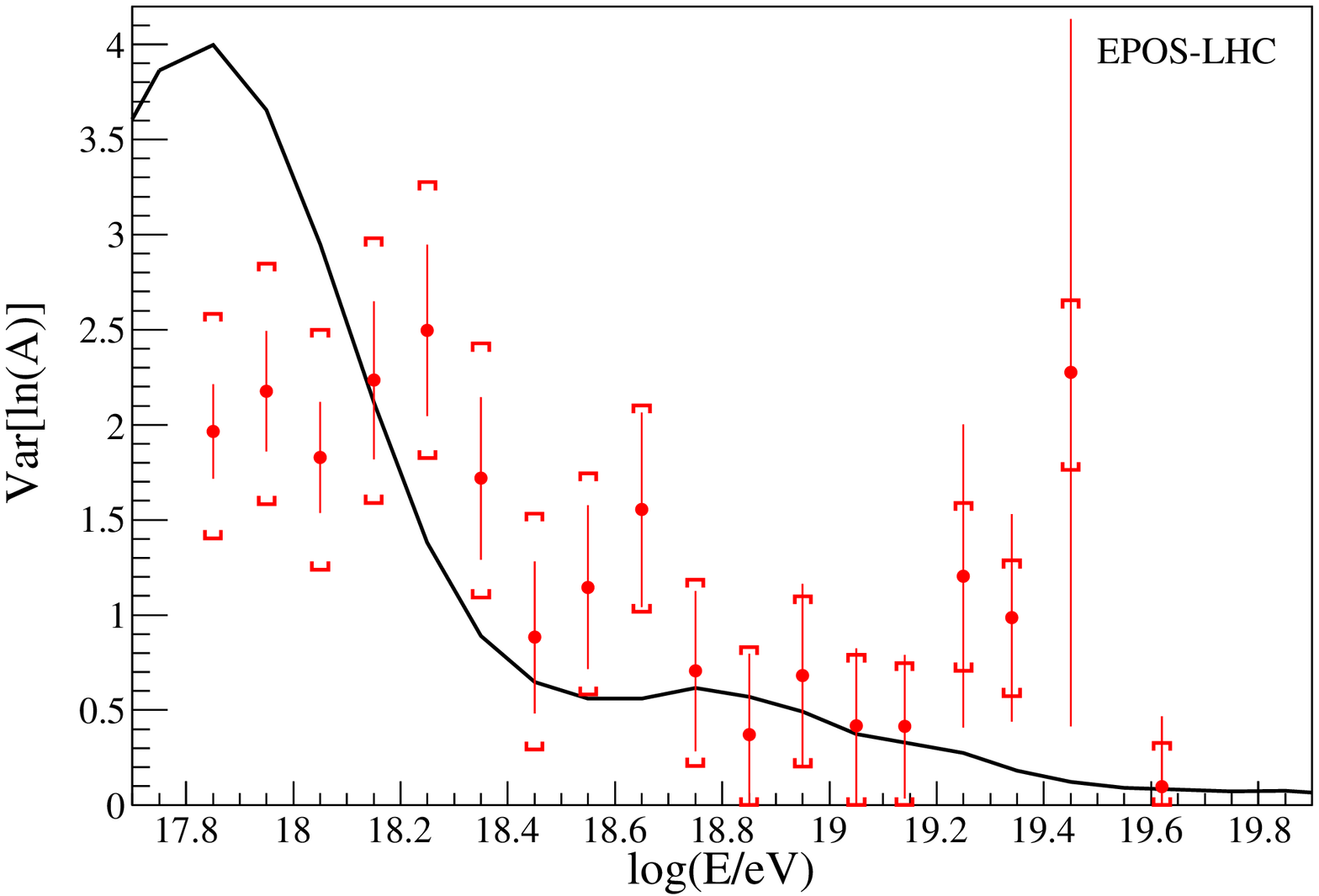}
\caption{Same as Fig.~\ref{M1}, but for $S(z)=1$. \label{M3}}
\end{figure}

In order to study the influence of the random magnetic field present in the source environment, a scenario 
with the same parameters as the previous one (with $S(z)=1$) but for $B=100$ G is considered. As can be seen 
from Fig.~\ref{M3}, also in this case a good fit of the spectrum is obtained. In this scenario, the injected 
spectrum is dominated by silicon and iron nuclei, and as can be seen from the top panel of the figure, the 
contribution of the other components is negligible. From the middle panel of the figure, it can be seen that 
$\langle \ln(A) \rangle$ is smaller than the one corresponding to the previous scenario. This is due to the 
fact that increasing the magnetic field intensity increases the number of nuclei that propagate diffusively 
through the source environment (see Fig.~\ref{SILengths}). The particles that propagate in the diffusive 
regime travel larger path lengths, which causes an increase of the number of interactions, mainly 
photodisintegrations, undergone by them.     

Photons and neutrinos are produced as a consequence of the UHECR propagation through the Universe.
These secondary particles are generated by the decay of pions produced by the photopion production
process undergone by the nuclei that interact with the low energy photons of the CMB and EBL. There
is also a contribution to the neutrino component that comes from neutron decay. The only energy loss 
undergone by the neutrinos is the one corresponding to the adiabatic expansion of the Universe. Unlike 
what happens to the neutrinos, the high energy photons interact with the low energy photons of the CMB 
and EBL, initiating electromagnetic cascades that develop in the intergalactic medium. As a result, the 
photon flux at Earth spans from the ultrahigh energy region down to energies below 1 GeV. Therefore, 
the UHECRs can contribute to the low energy diffuse photon background. 

\begin{figure}[!ht]
\includegraphics[width=8.5cm]{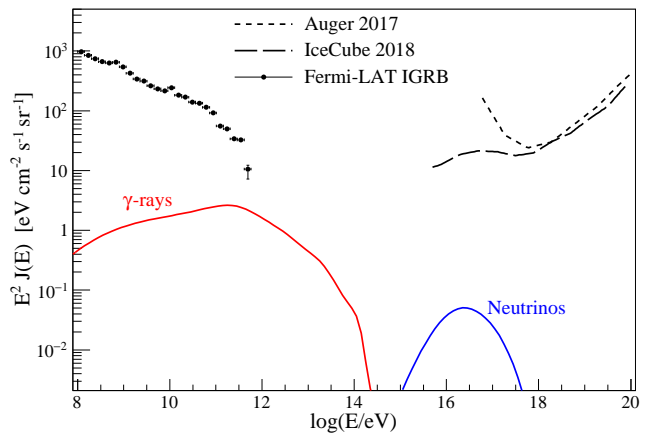}
\caption{$\gamma$-ray and all flavors neutrino flux expected for the model of Fig.~\ref{M3}. The data points
correspond to the IGRB obtained by \emph{Fermi}-LAT \cite{FermiLATIGRB:15}. Also shown are the upper limits 
to the neutrino flux at 90\% C.L., obtained by IceCube \cite{IceCube:18} and by Auger \cite{AugerNu:17}. 
\label{GammaNu}}
\end{figure}

The photon and neutrino fluxes corresponding to the two models compatible with the Auger data are 
calculated by using CRPropa 3. Figure \ref{GammaNu} shows the $\gamma$-ray and neutrino fluxes for the 
model corresponding to Fig.~\ref{M3} ($B=100$ G). As can be seen from the plot, the photon flux is 
smaller than the isotropic $\gamma$-ray background (IGRB) observed 
by the \emph{Fermi} Large Aea Telescope (\emph{Fermi}-LAT) \cite{FermiLATIGRB:15}. Moreover, the integral 
of the flux between 50 GeV and 2 TeV is $\sim 20$ times smaller than the 90\% C.L.~upper 
limit of Ref.~\cite{Supanitsky:16}, which was obtained by using the \emph{Fermi}-LAT analysis reported in 
Ref.~\cite{FermiLATPRL:16}. Also, from the figure it can be seen that the secondary neutrino flux is much 
smaller than the upper limits obtained by IceCube \cite{IceCube:18} and by Auger \cite{AugerNu:17}. Similar 
results are obtained for the model of Fig.~\ref{M2} ($B=1$ G). Low values of the $\gamma$-ray and neutrino 
fluxes are expected because it is very well known that the production of secondary particles, in models in 
which the high energy part of the cosmic ray flux is dominated by heavy nuclei, is much smaller than in 
the ones dominated by protons \cite{Aloisio:11}, which are still compatible with the neutrinos and 
$\gamma$-ray constraints in a region of the parametric space \cite{Supanitsky:16,Berezinsky:16}.

Increasing the luminosity of the photon gas present in the source environment makes the composition 
lighter; this is due to the decrease of the interaction lengths of the nuclei. In the limit of 
$\lambda_{PD},\, \lambda_{PP} \ll R$ all nuclear species are disintegrated before leaving the sphere, 
and then a light composition formed by protons and neutrons is obtained. It is found that models with 
$L\geq5\times10^{41}$ erg s$^{-1}$ are not compatible with the experimental data. Therefore, preferred 
models are such that $L \lesssim 10^{41}$ erg s$^{-1}$, which corresponds to low luminosity AGN (LLAGN) 
\cite{Ho:97,Ho:99}. It is worth noting that the central regions of these types of galaxies have been 
proposed as sources of the astrophysical neutrino flux observed by IceCube \cite{IceCubeNF:13}. In these 
models the high energy neutrinos are produced as a by-product of accelerated protons 
\cite{Kimura:15,Khiali:16}.    

The redshift evolution of the sources is poorly known. In general, the source evolution of AGN is assumed 
to increase very fast between $z=0$ and $z=1-2$ (like in Eq.~(\ref{Sz})) \cite{Hasinger:05}. However, in 
Ref.~\cite{Kimura:15} a nonevolving luminosity function for LLAGN is assumed. This is the case of the 
scenarios developed in this work, which are compatible with the Auger data. The argument in 
Ref.~\cite{Kimura:15} for this assumption is that LLAGN are similar to BL Lac objects (they both have a 
faint disk component), which have a luminosity function consistent with no evolution \cite{Ajello:14}.

It is worth mentioning that it is possible to fit the Auger spectrum assuming the evolution function
of Ref.~\cite{Kachel:17}, which corresponds to AGNs of $\log(L_X/\textrm{erg})=43.5$. However,
the composition predicted in this case is heavier at high energies ($\gtrsim 10^{18.8}$ eV) and 
lighter at low energies ($\lesssim 10^{18.3}$ eV) than the one obtained by using EPOS-LHC to analyze 
the Auger data. The behavior of the high energy part of the composition profile is consistent with the 
results obtained in Ref.~\cite{Kachel:17}. Therefore, also in this case the interaction of the nuclei 
with the protons present in a second region, surrounding the photon gas and filled with a proton gas, 
would be an appropriated mechanism to obtain a lighter composition at high energies, as it proposed in 
Ref.~\cite{Kachel:17}. 

In Ref.~\cite{Kachel:17}, a one-dimensional approach for the propagation of the nuclei in the source 
environment is considered. In that approximation, only the diffusive regime of propagation is taken 
into account. The escape time used in these types of calculations is taken as 
$\tau(E) =\tau_0\, [E/(Z\, E_0)]^{-\delta}$, where $\tau_0$ is a normalization constant, $E$ is the 
energy of the nucleus, $Z$ is its charge number, $E_0$ is a reference energy, and $\delta$ is a 
positive index. In the leaky box model approximation $\tau(E) \propto 1/D(E)$, where $D(E)$ is the 
diffusion coefficient. Therefore, the index $\delta$ gives the energy dependence of the diffusion 
coefficient. The escape time in these types of models is a decreasing function of the energy, which 
is valid up to a distance of the order of the size of the source environment region. Therefore, 
depending on the parameters used for the escape time in the one-dimensional calculation and the size of 
the source environment used in the three-dimensional approach, the high energy nuclei can escape from 
the source environment region before, compared to the case in which the particle propagates ballistically. 
In this case, a larger light component is expected at low energies for the three-dimensional calculation 
due to the larger number of photodisintegrations undergone by the high energy nuclei. This is the case 
for the model in which the source evolution function of Ref.~\cite{Kachel:17} is considered. The larger 
number of light nuclei at low energies makes the composition lighter than the one obtained by Auger, using 
EPOS-LHC to interpret the data, and also a harder spectral index is required, $\Gamma=1$, to obtain a 
good fit of the spectrum compared to the one considered in Ref.~\cite{Kachel:17}.       

It should be noted that an independent composition analysis in the region of the spectrum below the
ankle will be possible, in the near future, by using the information of the muon detectors of AMIGA 
(Auger Muons and Infill for the Ground Array) that are being installed at the Auger site 
\cite{Figueira:17}. As mentioned before, the muon content of the shower is very sensitive to the nature 
of the primary cosmic ray.

\section{Conclusions}

In this work, we have studied the possibility that the presumed extragalactic light component that dominates
the UHECR flux below the ankle originates from the photodisintegration of more energetic and heavier nuclei in the 
photon gas present in the central regions of active galaxies. In this scenario, the UHECRs are accelerated near 
the supermassive black hole present in the central region of these galaxies. Note that these types of models 
require only one population of UHECR sources to explain the experimental data above $\sim 10^{18}$ eV. 

We have found that low luminosity active galaxies with no source evolution are compatible with present composition 
and flux data, within the systematic uncertainties introduced by the high energy hadronic interaction models. It is 
worth mentioning that these types of astronomical objects have been proposed as the source of the high energy 
neutrinos observed by IceCube. We have also found that increasing the intensity of the random magnetic field in 
the source environment makes the composition observed at Earth lighter, as expected. However, we have proved that 
models with larger values of luminosity of the photon gas or with a strong source evolution are incompatible with 
present experimental data.

\begin{acknowledgments}

A.~D.~S.~and A.~E.~are member of the Carrera del Investigador Cient\'ifico of CONICET, Argentina. This work is 
supported by ANPCyT PICT-2015-2752, Argentina. The authors thank the members of the Pierre Auger Collaboration 
for useful discussions and R. Clay and C. Dobrigkeit Chinellato for reviewing the manuscript.

\end{acknowledgments}

\end{document}